\documentclass[aps,amsmath,showpacs,amsfonts,11pt]{revtex4}
\usepackage{graphicx}
\usepackage{epstopdf}
\usepackage{epsfig,graphicx}
\usepackage[english]{babel}
\usepackage{amsfonts}
\usepackage{amsmath}
\usepackage{latexsym}
\usepackage{graphics,bm}
\usepackage{dcolumn}
\usepackage{bm}
\usepackage{rotating}
\begin{document}

\title{Controllable optical bistability and Fano line shape in a hybrid optomechanical system assisted by Kerr medium: Possibility of all optical switching}

\author{Aranya B Bhattacherjee$^{1}$ and Muhammad S. Hasan$^{2}$}

\affiliation{$^{1}$ School of Physical Sciences, Jawaharlal Nehru University, New Delhi-110067, India  \\
 $^{2}$ Department of Physics,National Institute of Technology, Calicut, India}

\begin{abstract}
We theoretically analyze the optical and optomechanical nonlinearity present in a hybrid system consisting of a quantum dot(QD) coupled to an optomechanical cavity in the presence of a nonlinear Kerr medium, and show that this hybrid system can be used as an all optical switch. A high degree of control and tunability via the QD-cavity coupling strength, the Kerr and the optomechanical nonlinearity over the bistable behavior shown by the mean intracavity optical field and the power transmission of the weak probe field can be achieved.The results obtained in this investigation has the potential to be used for designing efficient all-optical switch and high sensitive sensors for use in Telecom systems.

\end{abstract}

\maketitle

\section{Introduction}

One of the obstacles of Moore's law imposed by device miniaturization is the technological development of classical computing reaching the fundamental limit. However this limitation can be overcome by integrating quantum optics on existing computing platforms such as solid state quantum systems \citep{Vu1,jel2,mich3,mas4}. In particular quantum dots (QDs) are considered highly attractive candidates for such applications due to their optical tunability, their narrow linewidths and potential for implementation of optoelectronic devices. To make such applications a reality, complete coherent control of the quantum device is necessary. One such quantum device is an all optical switch with a single QD strongly coupled to a nanocavity \citep{maj5,eng6,bos7}. By confining the QD in a cavity, repeated interactions with the trapped cavity photons leads to the strong coupling regime if the interaction strength overcomes the losses \citep{lau8,dor9}. In recent years, several experiments have demonstrated efficient out-coupling using strongly coupled QD-cavity systems \citep{eng10,bro11,rei12,kim13}. The QD-cavity systems has also proved to be extremely versatile since it is possible to control the spontaneous emission of the QD in the optical cavity \citep{lod14}.

Cavity optomechanics is a rapidly growing field of research in which a coherent coupling between the cavity optical modes and the mechanical modes of the oscillator can be achieved via the radiation pressure exerted by the trapped cavity photons \citep{asp15,asp16,asp17,mar18,kip19}. Rapid technological advancements in this field has led to marked achievements such as ultrahigh-precision measurements \citep{teu20}, gravitational wave detectors \citep{arv21}, quantum information processing \citep{wan22,don23}, quantum entanglement \citep{hof24,wan25,agg26} and optomechanically induced transparency (OMIT) \citep{wei27,ma28,aga29,tas30,saf31,ma32,jia33,zha34,liu35}. In optomechanical systems, a high degree of nonlinearity exists between the optical field and mechanical mode, which gives rise to optical bistability and multistability \citep{zha34,cha36,gho37,kyr38}. This optical phenomena has practical applications in all optical switching \citep{jia39,sar40} and memory storage \citep{yan41,gao42}. Optomechanical systems also exhibits an interesting optical phenomena called Fano resonance \citep{fan43} which is based on quantum coherence and interference. Fano resonance is characterized by a sharp asymmetric line profile. Fano resonance is different from electromagnetically induced transparency (EIT) and OMIT, both of which have symmetric line profile. The asymmetry of the Fano line shape and enhanced interference effect has attracted many theoretical \citep{lon44,yas45,qu46,lu47} as well as experimental investigations \citep{li48,hay49,lei50,mor51}. EIT is a quantum interference effect arising from different transition pathways of optical fields \citep{liu35}. In the EIT effect, an abnormal dispersion occurs with the opening of a transparency window, resulting in slow light i.e reduction of light group velocity \citep{kas52,hau53}. This phenomenon implies that light can be stored in atomic ensembles \citep{phi54,liu55}. Slow light has important applications in optical networks \citep{boy56}, quantum networks \citep{kim57} and quantum memory \citep{luk58,lvo59}.

In this work, we present a complete analytical investigation based on a hybrid optomechanical system consisting of an optomechanical cavity containing a QD and a nonlinear Kerr medium. In particular, we study the bistable behaviour shown by our proposed hybrid optomechanical system and calculate the power transmission of a weak probe field under the action of a strong pump field. We clearly show a high degree of control and tunability can be achieved over the generated optical bistability and Fano resonance  using the QD-cavity coupling strength, Kerr and optomechanical nonlinearity.

\section{Theoretical Model}

\begin{figure}[h]
\hspace{-0.0cm}
\includegraphics [scale=0.30]{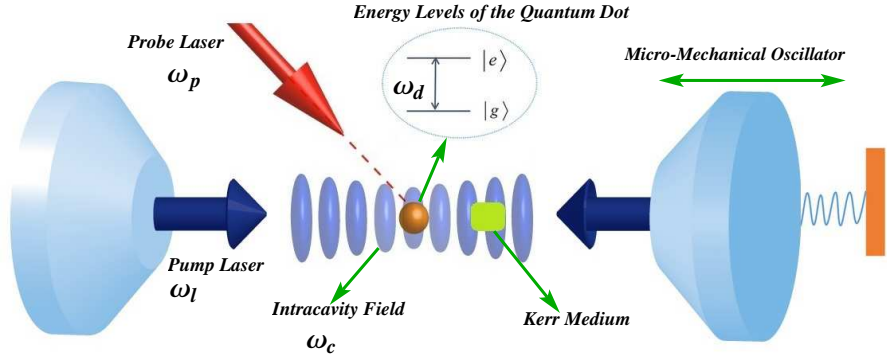}\\
\caption{Schematic diagram of the cavity optomechanical system consisting of a quantum dot coupled to the cavity mode in the presence of a nonlinear Kerr medium. The cavity optomechanical system consists of  fixed mirror and a movable mirror. The cavity is driven by a strong pump laser and the system is probed by a weak laser. }
\label{Fig.1}
\end{figure}

We consider a hybrid quantum system consisting of an optical cavity and optomechanics, where a two-level quantum dot is coupled to a confined photon mode and a micro-mechanical mode via radiation pressure in the presence of a Kerr nonlinear medium. The physical setup and the energy level configuration of the QD are shown in \ref{Fig.1}. The proposed hybrid model has two distinct non-linearities. Firstly the optical non-linearity is introduced by the Kerr medium. Secondly, the system has an optomechanical non-linearity due to the coupling of the single optical mode to the micro-mechanical resonator. The optomechanical coupling strength between the optical mode and the mechanical mode is characterized by $G$. The coupling between the optical mode and the QD is described by the Jaynes-Cummings type of interaction characterized by the parameter $g$.
We assume that the system is coherently driven by external input laser fields consisting of a strong pump field of frequency $\omega_{l}$ and a weak probe field of frequency $\omega_{p}$ along the cavity axis. The external input laser field is denoted by $a_{in}(t)=E_{l} e^{-i\omega_{l}t}+E_{p} e^{-i\omega_{p}t}$ with field strengths $E_{l}$ and $E_{p}$ of the pump and probe lasers respectively. The field strengths $E_{l}$ and $E_{p}$ are related to the laser powers as $E_{l}=\sqrt{P_{l}/\hbar \omega_{l}}$ and $E_{p}=\sqrt{P_{p}/\hbar \omega_{p}}$, where $P_{l}$ and $P_{p}$ are the powers of the pump field and the probe field respectively. We assume without loss of generality that $E_{l}$ and $E_{p}$ are real. Experimentally $E_{l}>> E_{p}$.

The Hamiltonian of the whole system is given by

\begin{eqnarray}
H &=& \frac{p^{2}}{2m}+\frac{1}{2} m \omega_{m}^{2} q^{2}+ \hbar \omega_{c} a^{\dagger}a +\hbar \omega_{d} \sigma^{\dagger} \sigma\\ \nonumber
&-&   \frac{\hbar}{2} \kappa a^{\dagger} a^{\dagger}a a + \hbar g \left( a^{\dagger} \sigma + \sigma^{\dagger} a \right)-\hbar G a^{\dagger}a q \\ \nonumber
&+& i \hbar \sqrt{2 \eta} \left( a^{\dagger} E_{l} e^{-i \omega_{l}t}-a E_{l}^{*}  e^{i \omega_{l}t}\right) \\ \nonumber
 &+& i \hbar \sqrt{2 \eta} \left( a^{\dagger} E_{p} e^{-i \omega_{p}t}-a E_{p}^{*}  e^{i \omega_{p}t}\right)
\end{eqnarray}

The first and the second terms give the energy of the mechanical oscillator, where $q$ and $p$ are the position and momentum operators of the mechanical oscillator with effective mass $m$ and natural frequency $\omega_{m}$, satisfying the commutation relation $[q,p]=i \hbar$.  The third term represents the free energy of the cavity mode with frequency $\omega_{c}$ and photon creation (annihilation) operators $a^{\dagger}$ ($a$). The fourth term is the energy of the two level semiconductor QD with $\hbar \omega_{d}$ as the energy difference between the excited and the ground level. The raising and lowering operators of the two-level QD with transition frequency $\omega_{d}$ is $\sigma^{\dagger}$ and $\sigma$  respectively. The operators $\sigma^{\dagger}$ and $\sigma$ are expressed in terms of the usual Pauli spin operators $\sigma_{x,y,z}$ as:
$\sigma^{\dagger}=\frac{\sigma_{x}+i \sigma_{y}}{2}$ and $\sigma= \frac{\sigma_{x}-i \sigma_{y}}{2}$. The fifth term describes the Kerr non-linearity with non-linear coefficient $\kappa$. Giant optical Kerr nonlinearities are obtained by placing a $\zeta^{(3)}$ medium inside a cavity \citep{Imm}, with $\kappa = 3 \omega_{c}^{2} Re[\zeta^{(3)}]/2 \epsilon_{o} V_{c}$, $\epsilon_{o}$ being the dielectric constant of the medium, $V_{c}$ being the volume of the cavity, and $\zeta^{(3)}$ being the third-order nonlinear susceptibility.  The sixth term describes the QD-field interaction with coupling $g$. The seventh term shows the interaction between the cavity field and the mechanical oscillator with optomechanical coupling strength $G=\frac{\omega_{c}}{L}\sqrt{\frac{\hbar}{m \omega_{m}}}$, which is determined by the shift of the cavity resonance per displacement of the mechanical oscillator. Here $L$ is the length of the optical cavity. Finally the eighth and the ninth terms represents the effect of the strong pump laser and the weak probe laser respectively.
In a frame rotating at the driving laser frequency $\omega_{l}$, the Hamiltonian can then be written as,

\begin{eqnarray}
H &=& \frac{p^{2}}{2m}+\frac{1}{2} m \omega_{m}^{2} q^{2}+ \hbar \Delta_{c} a^{\dagger}a +\hbar \Delta_{d} \sigma^{\dagger} \sigma  \\ \nonumber
 &-&  \frac{\hbar}{2} \kappa a^{\dagger} a^{\dagger}a a + \hbar g \left( a^{\dagger} \sigma + \sigma^{\dagger} a \right)-\hbar G a^{\dagger}a q  \\ \nonumber
 &+&  i \hbar \sqrt{2 \eta} E_{l} \left( a^{\dagger} -a \right) + i \hbar \sqrt{2 \eta} E_{p} \left( a^{\dagger}  e^{-i \Delta_{p}t}-a   e^{i \Delta_{p}t}\right),
\end{eqnarray}

where $\Delta_{c}= \omega_{c}-\omega_{l}$, $\Delta_{d}= \omega_{d}-\omega_{l}$ and $\Delta_{p}= \omega_{p}-\omega_{l}$ are the detunings of the cavity field frequency, the QD transition frequency and the probe field frequency respectively. The factors described above may be used to coherently control the optical bistability and the power transmission.

\section{Controlled optical bistability}

In this section, we will show how optical bistability for the cavity mode can be modified and controlled by the various system parameters. Before proceeding ahead, we first define the dimensionless position and momentum operators $Q$ and $P$ as $Q=q \sqrt{\frac{m \omega_{m}}{\hbar}}$ and $P=p \sqrt{\frac{1}{\hbar m \omega_{m}}}$ for the mechanical oscillator. Now we write down the Heisenberg-Langevin equations of motion based on the effective Hamiltonian of Eqn.(2) as:

\begin{equation}
\frac{dQ}{dt}= \omega_{m} P,
\end{equation}

\begin{equation}
\frac{dP}{dt}=-\omega_{m} Q+ \chi \omega_{m} a^{\dagger}a-\gamma_{m} P,
\end{equation}

\begin{eqnarray}
\frac{da}{dt} &=& -i \Delta_{c} a +i \chi \omega_{m} a Q-i g \sigma+i \kappa a^{\dagger}aa+ \sqrt{2 \eta} E_{l} \\ \nonumber
&+& \sqrt{2 \eta} E_{p} e^{-i \Delta_{p}t}- \eta a,
\end{eqnarray}

\begin{equation}
\frac{d \sigma}{dt}= i \Delta_{d} \sigma \sigma_{z}+i g a \sigma_{z}-\gamma_{a} \sigma,
\end{equation}

where $\eta$, $\gamma_{a}$ and $\gamma_{m}$ are the decay rates associated with the cavity mode, the QD and the mechanical mode respectively. Here $\chi=\frac{G}{\omega_{m}}\sqrt{\frac{\hbar}{m \omega_{m}}}$ is the scaled opto-mechanical coupling constant. The steady state of the system is analyzed by reducing the operators to their mean values and dropping the quantum noise terms since they have zero mean values.The steady state of the system is characterized by large amplitude semiclassical solutions for all the three degrees of freedom i.e the optical field, mechanical mode and the QD. Also for the case when the driving field is much stronger than the probe field i.e $E_{l}>> E_{p}$, the weak probe field is treated as noise. The Kerr and the optomechanical nonlinearity inherent in the equations of motion of our system indicates the presence of bistability. In the mean field approximation $<Q a> = <Q> <a>$ and hence the steady state solutions of Eqns.(3)-(6) can be obtained as,

\begin{equation}
Q_{s}=\chi |a_{s}|^{2},
\end{equation}

\begin{equation}
P_{s}=0,
\end{equation}

\begin{equation}
\sigma_{s}= - \frac{i g a_{s} <\sigma_{z}>_{s}}{(\gamma_{a}+i \Delta_{d})},
\end{equation}

\begin{equation}
a_{s}= \frac{\sqrt{2 \eta }E_{l}}{\left\{ i \Delta_{c}+ \frac{g^2 <\sigma_{z}>_{s}}{\gamma_{a}^2+ \Delta_{d}^{2}}(\gamma_{a}-i \Delta_{d})+\eta-i |a_{s}|^{2} (\omega_{m} \chi^2+\kappa) \right\}}.
\end{equation}

\begin{figure}[t]
\hspace{-0.0cm}
\begin{tabular}{cc}
\includegraphics [scale=0.60]{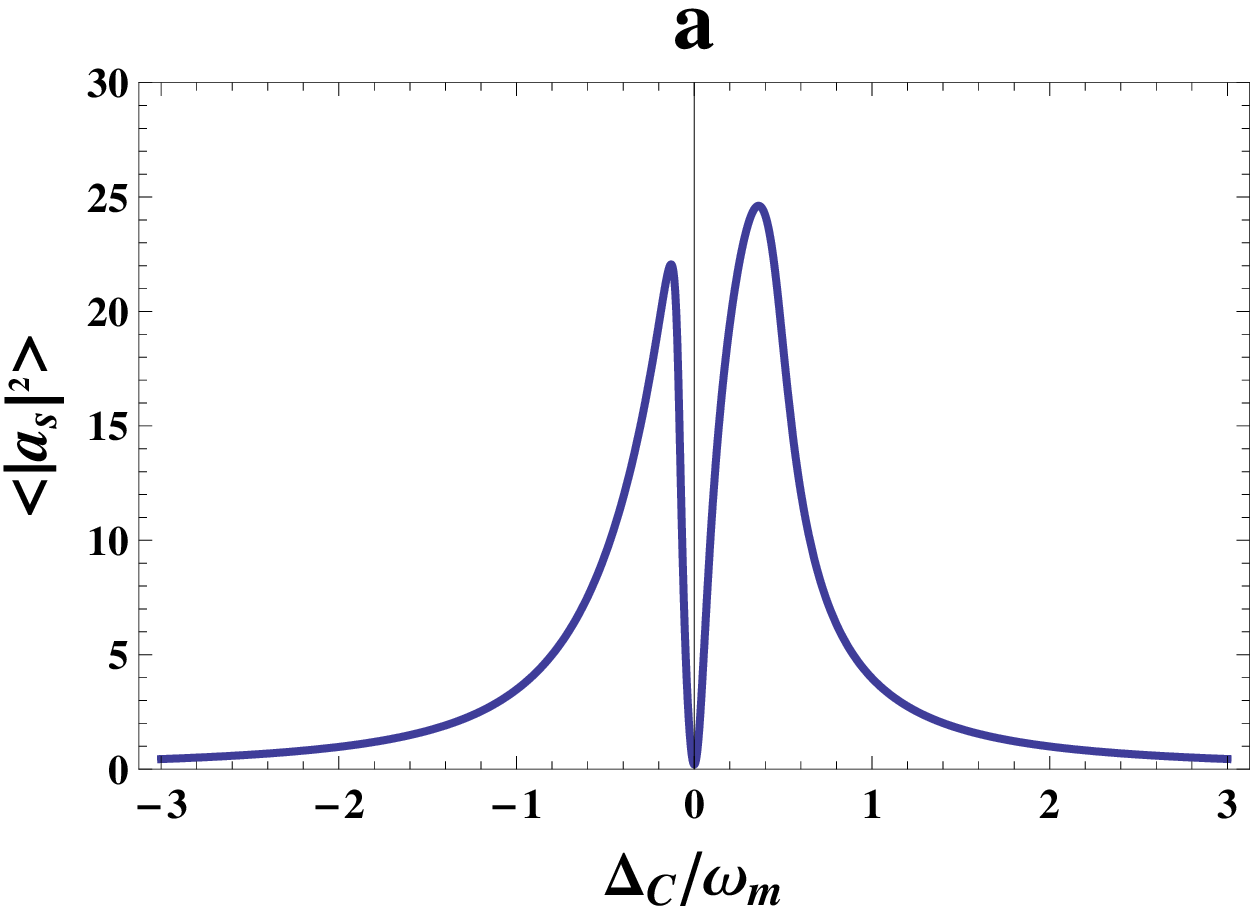}& \includegraphics [scale=0.60]{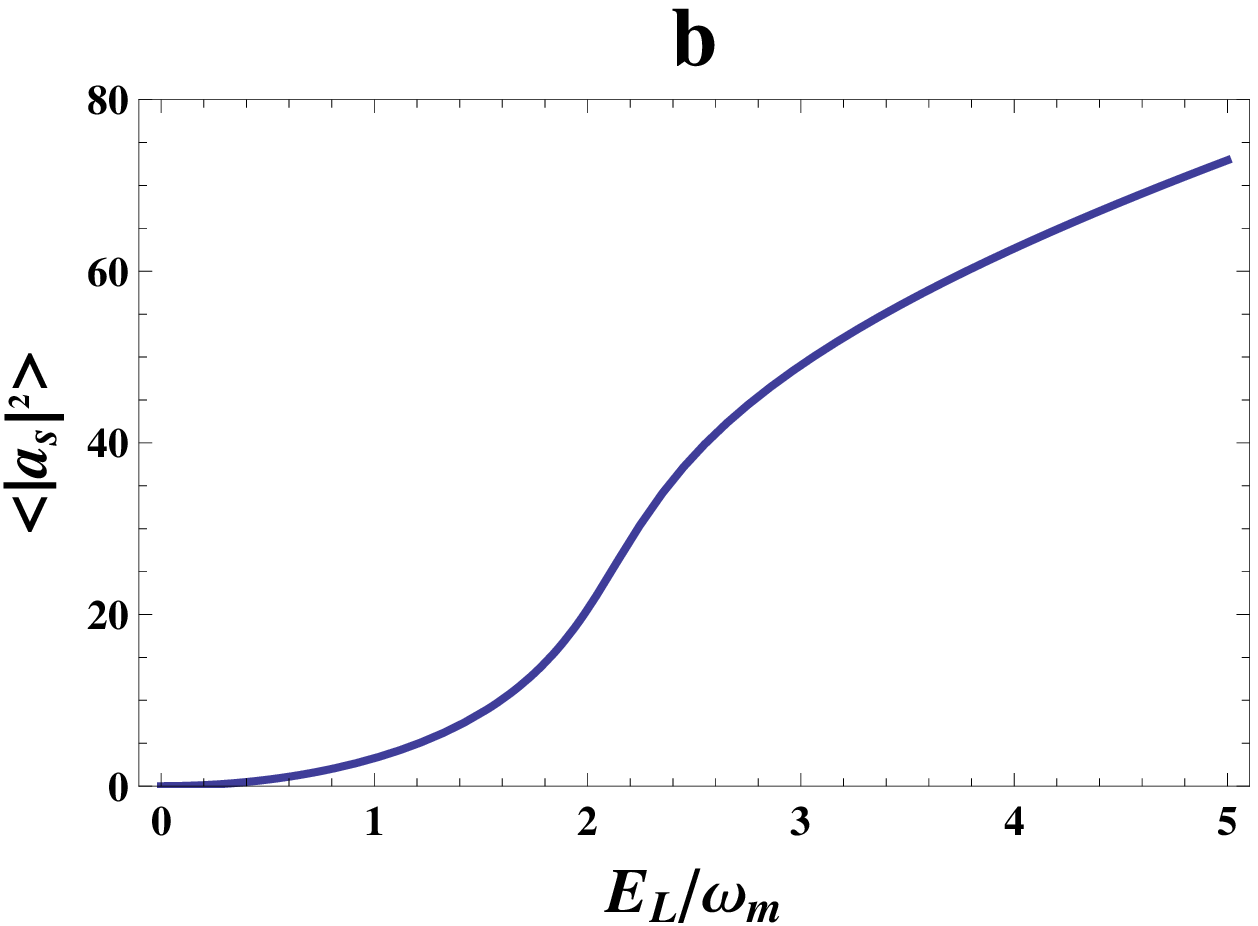}\\
\end{tabular}
\caption{(a) Plot of mean intracavity photon number $|a_{s}|^{2}$ versus $\Delta_{c}/\omega_{m}$. (b) Plot of mean intracavity photon number $|a_{s}|^{2}$ versus the dimensionless input pump field strength $E_{l}/\omega_{m}$. Other system parameters used are $\Delta_{d}=\Delta_{c}$, $\eta=0.4 \omega_{m}$, $E_{l}=2 \omega_{m}$, $\gamma_{a}=0.01 \omega_{m}$, $\kappa=0.01 \omega_{m}$, $\chi=0.04$ and $g=0.2 \omega_{m}$. In plot (b), $\Delta_{c}=0.5 \omega_{m}$.}
\label{Fig.2}
\end{figure}

\begin{figure}[t]
\hspace{-0.0cm}
\begin{tabular}{cc}
\includegraphics [scale=0.60]{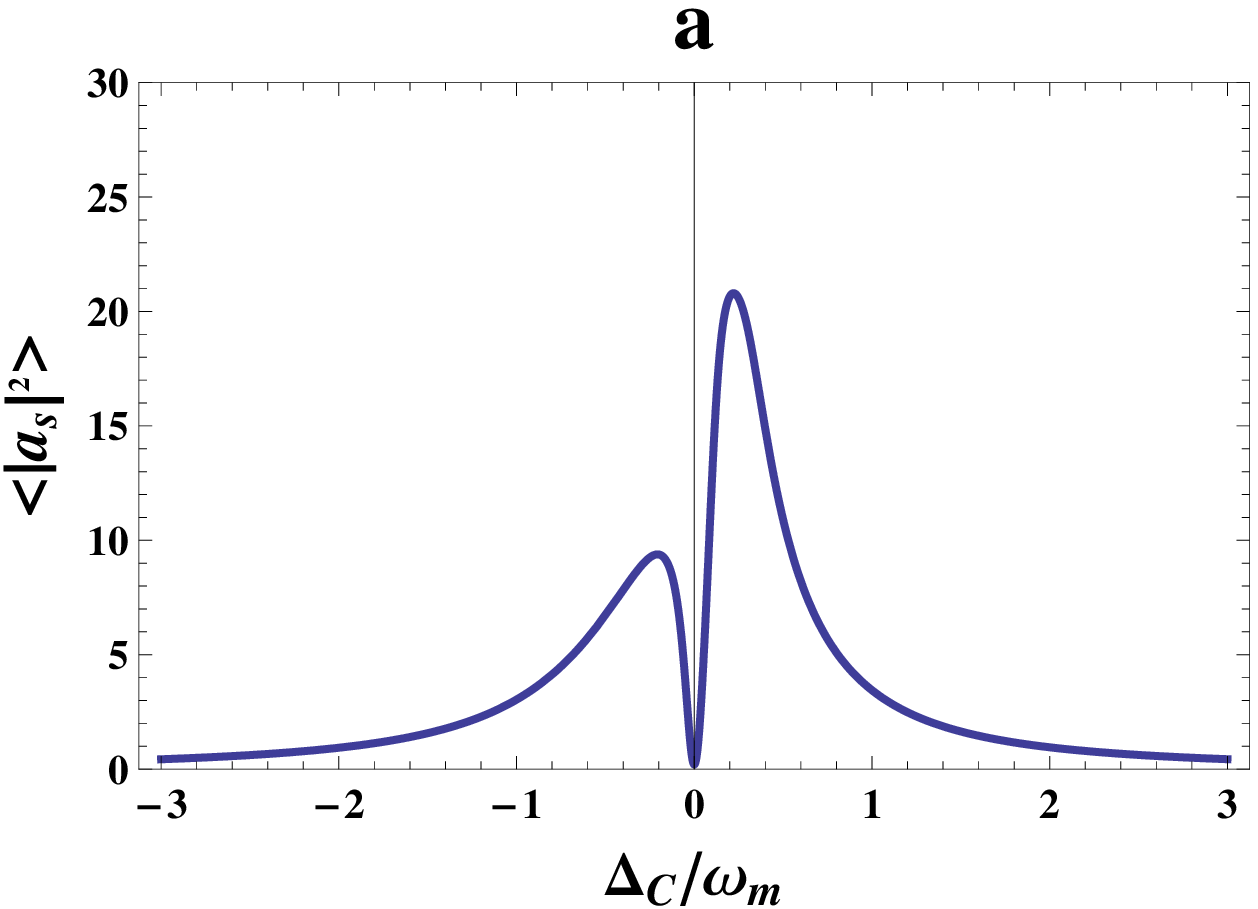}& \includegraphics [scale=0.60]{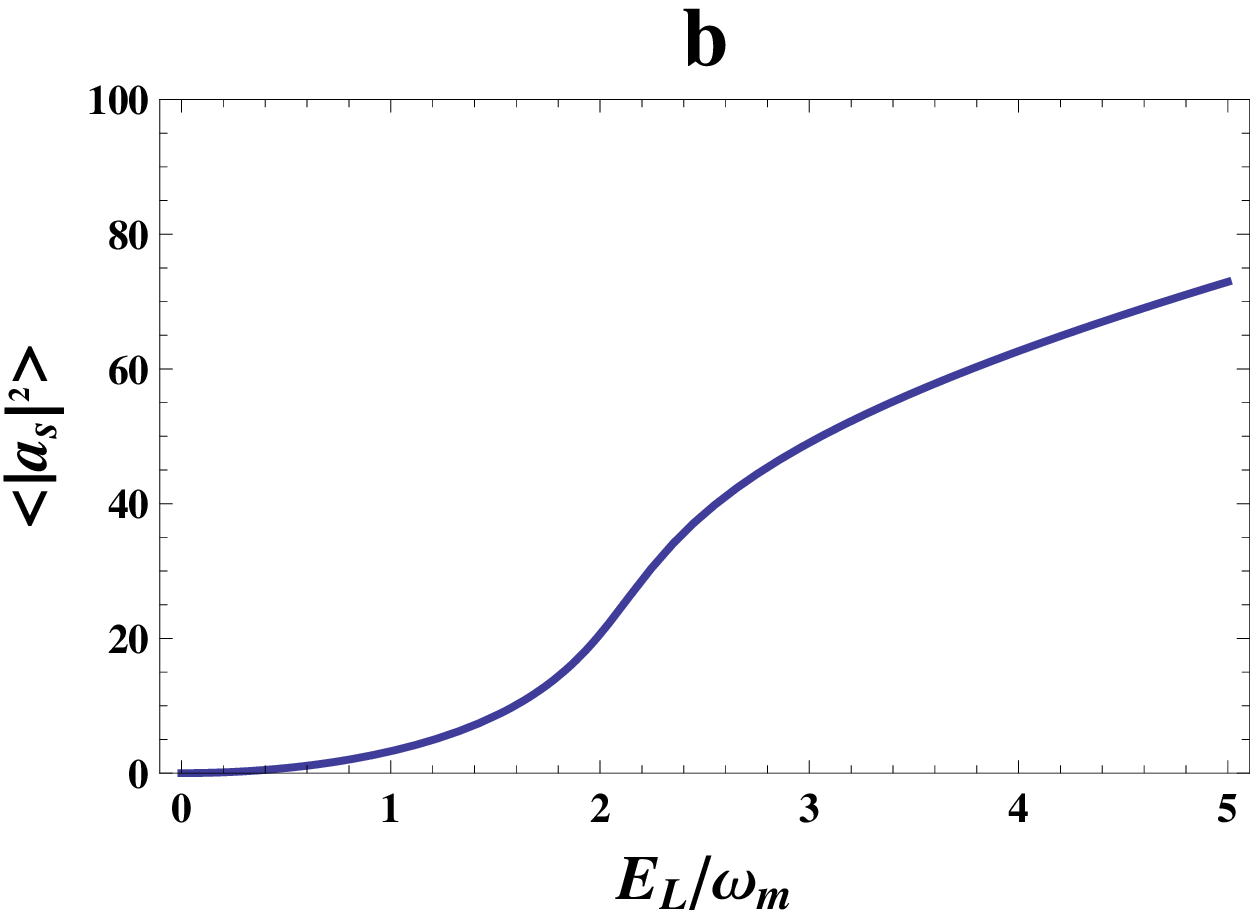}\\
\end{tabular}
\caption{(a) Plot of mean intracavity photon number $|a_{s}|^{2}$ versus $\Delta_{c}/\omega_{m}$. (b) Plot of mean intracavity photon number $|a_{s}|^{2}$ versus the dimensionless input pump field strength $E_{l}/\omega_{m}$. Here $\Delta_{d}= - \Delta_{c}$ and other parameters take the same values as in Fig.2.  In plot (b) $\Delta_{c}=0.5 \omega_{m}$.}
\label{Fig.3}
\end{figure}

In the low-excitation regime the steady state value $<\sigma_{z}>_{s}=-1$ since the QD stays only in the ground state. This neglects the nonlinear nature of the QD. From Eqn.(10), we derive the following expression for $|a_{s}|^{2}$ which indicates the existence of bistable behaviour for certain values of the system parameters,

\begin{eqnarray}
&& |a_{s}|^{6} (\omega_{m} \chi^{2} + \kappa)^{2} \\ \nonumber
&+& |a_{s}|^{4}(\omega_{m} \chi^{2}+\kappa)\left(  \frac{g^2 \Delta_{d}}{\gamma_{a}^2+ \Delta_{d}^{2}}-\Delta_{c}\right) \\ \nonumber
&+& |a_{s}|^{2} \left( \eta^{2} +\Delta_{c}^{2}+ \frac{g^{4}+2 \eta \gamma_{a} g^{2}-2 \Delta_{d} \Delta_{c} g^{2}}{\gamma_{a}^{2}+\Delta_{d}^{2}} \right)= \tilde{E_{l}^{2}},
\end{eqnarray}

where $\tilde{E_{l}}= \sqrt{2 \eta} E_{l}$. For bistability to exist, one must have $\frac{\partial \tilde{E_{l}^{2}}}{\partial |a_{s}|^{2}}=0$. This yields the following equation:

\begin{eqnarray}
&& 3 |a_{s}|^{4} (\omega_{m} \chi^{2} + \kappa)^{2}  \\ \nonumber
&+& 4 |a_{s}|^{2}(\omega_{m} \chi^{2}+\kappa)\left(  \frac{g^2 \Delta_{d}}{\gamma_{a}^2+ \Delta_{d}^{2}}-\Delta_{c}\right) \\ \nonumber
&+& \left( \eta^{2}+\Delta_{c}^{2}+ \frac{g^{4}+2 \eta \gamma_{a} g^{2}-2 \Delta_{d} \Delta_{c} g^{2}}{\gamma_{a}^{2}+\Delta_{d}^{2}} \right)= 0
\end{eqnarray}

Eqn.(12) is quadratic in $|a_{s}|^{2}$ which has two distinct roots if the discriminant is positive:

\begin{equation}
\frac{4 (\omega_{m} \chi^{2}+\kappa)^{2}}{(\gamma_{a}^{2}+\Delta_{d}^{2})^{2}} \left\{4 g^{4} \Delta_{d}^{2}+(\Delta_{c}^{2}-3 \eta^{2}) (\gamma_{a}^{2}+\Delta_{d}^{2})^{2}-(2 g^{2} \Delta_{c} \Delta_{d}+3 g^{4}+6 \eta \gamma_{a} g^{2})(\gamma_{a}^{2}+\Delta_{d}^{2})  \right\} >0.
\end{equation}

Clearly for weak Kerr nonlinearity $\kappa$ and weak optomechanical coupling strength $\chi$, bistability is absent since the discriminant is vanishingly small. In order to have appreciable bistable bahaviour, either $\kappa$ or $\chi$ should be large. We plot the stationary value for the intracavity photon number $|a_{s}|^{2}$ as a function of $\Delta_{c}/\omega_{m}$ for different values of the system parameters. Based on earlier experimental studies \citep{asp16, gro,gig,chak,reith}, the parameters vary in the following range: $L=10^{-3} - 25 \times 10^{-3} m$ , $m= 5 - 145 ng$, $\omega_{m}=1 - 10 MHz$, $\eta=0.2 - 0.5 \omega_{m}$, $\gamma_{m}=\omega_{m}/ Q$, where $Q$ is the quality factor of the cavity. To analyze the occurrence of bistability in the intra-cavity optical field in our proposed hybrid optomechanical system first we consider the case when both the Kerr and the optomechanical nonlinearity are small and comparable. Fig.2 and 3 are the plots corresponding to small Kerr ($\kappa=0.01 \omega_{m}$) and optomechanical ($\chi=0.04$) nonlinearity. It is easy to see that for small nonlinearity, bistable behaviour is absent. Interestingly both Figs. 2(a) and Fig. 3(a) exhibits the switching characteristics of the all-optical switch around $\Delta_{c}/\omega_{m}=0$. The asymmetric structure of the split resonance around $\Delta_{c}/\omega_{m}=0$ is due to the presence of Kerr nonlinearity. In the absence of the Kerr medium ($\kappa=0$), the symmetry in the split resonance is restored (figure not shown). Fig.2(b) and 3(b) shows the onset of nonlinear behaviour of $|a_{s}|^{2}$ as a function of $E_{l}/\omega_{m}$. Since both the nonlinearities are weak, the system fails to exhibit bistable behaviour.
When considering a strong kerr nonlinearity, $\kappa=0.1 \omega_{m}$ compared to optomechanical nonlinearity $\chi=0.04$, bistable behaviour appears as seen in Fig.4. Along with bistability, the system retains its switching characteristics around $\Delta_{c}/\omega_{m}=0$. For strong Kerr nonlinearity, Fig.4(b) exhibits the hysteresis curve for $|a_{s}|^{2}$ with respect to driving field strength $E_{l}/\omega_{m}$. Clearly starting from low value of $E_{l}$, as we increase the driving laser field strength, the intra-cavity intensity $|a_{s}|^{2}$ jumps from the lower stable branch to the upper stable branch. A similar transition from the upper stable branch to the lower stable branch occurs while decreasing the driving laser field strength. A similar bistable behaviour emerges when a strong optomechanical nonlinearity $\chi=0.3$ is introduced together with a weak Kerr nonlinearity $\kappa=0.01 \omega_{m}$ (figure not shown). These observations point to the fact that the origin of the bistable behaviour are the two inherent nonlinearities.
Finally for weak nonlinearities ($\kappa=0.01 \omega_{m}$, $\chi=0.04$) and strong QD-cavity mode coupling strength $g=0.8 \omega_{m}$, the intra-cavity intensity $|a_{s}|^{2}$ as a function of $\Delta_{c}/\omega_{m}$ (Fig.5) displays the asymmetric switching behaviour but with a much broader zero intensity window around $\Delta_{c}/\omega_{m}=0$. For a weak QD-cavity mode  coupling, the zero intensity window is much narrow as evident from Fig.2(a) and 3(a).

\begin{figure}[t]
\hspace{-0.0cm}
\begin{tabular}{cc}
\includegraphics [scale=0.60]{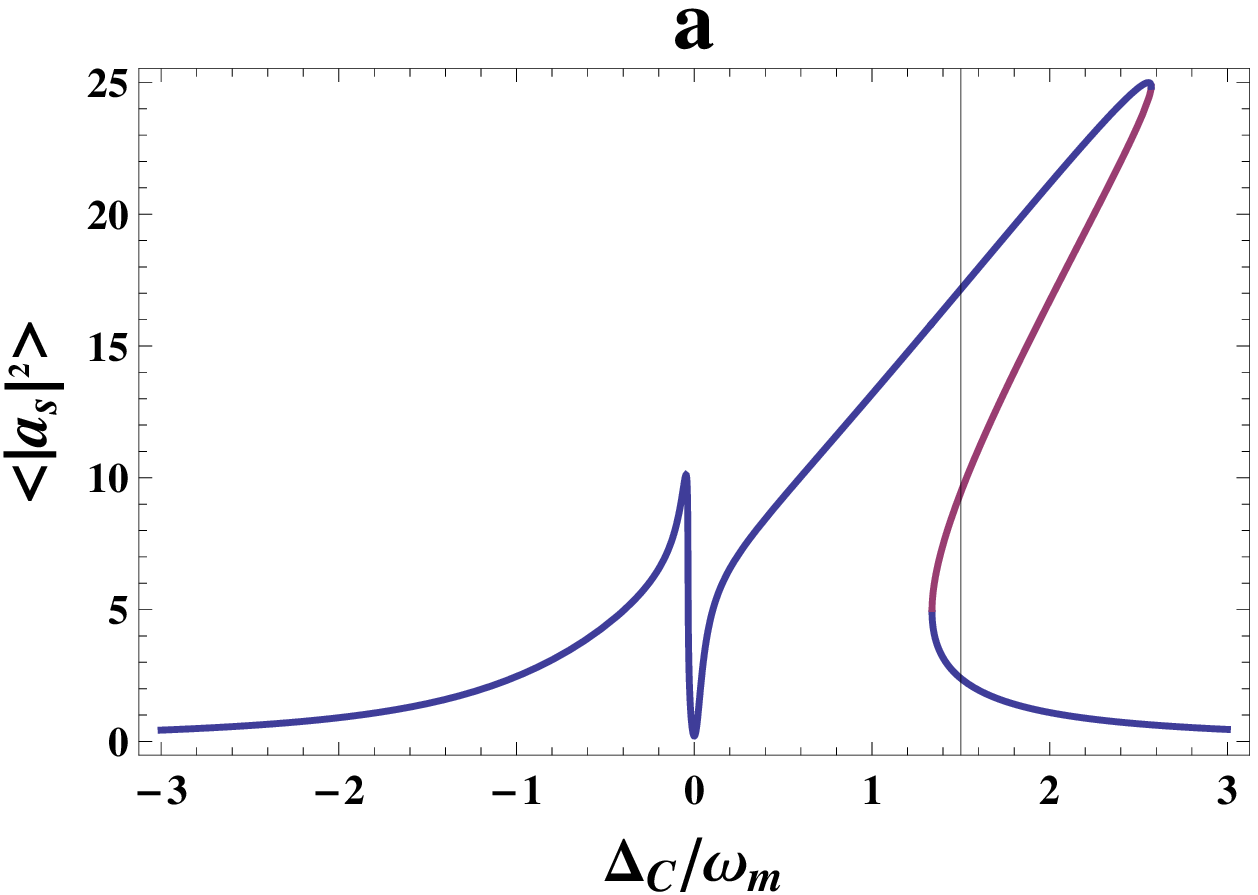}& \includegraphics [scale=0.60]{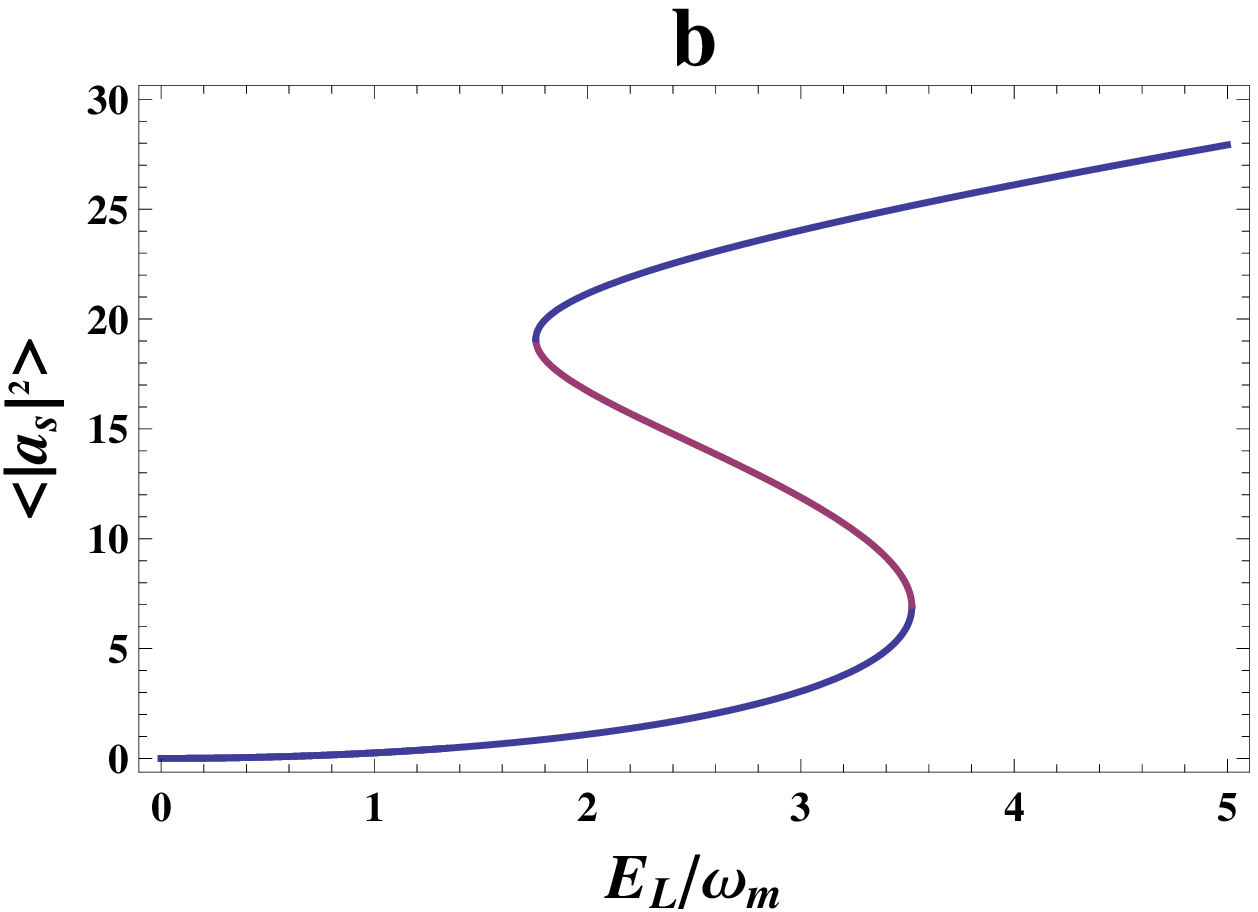}\\
\end{tabular}
\caption{(a) Plot of mean intracavity photon number $|a_{s}|^{2}$ versus $\Delta_{c}/\omega_{m}$. (b) Plot of mean intracavity photon number $|a_{s}|^{2}$ versus the dimensionless input pump field strength $E_{l}/\omega_{m}$. Here $\Delta_{d}=\Delta_{c}$ and $\kappa=0.1 \omega_{m}$. Other parameters are same as in Fig.2.}
\label{Fig.4}
\end{figure}

From the above discussions, we can arrive at the conclusion that the generated bistability is closely related to the Kerr and optomechanical nonlinearities. The hybrid optomechanics considered here enables a controlled bistable switching of the intra-cavity intensity by appropriately adjusting the Kerr nonlinearity $\kappa$ and the optomechanical coupling $\chi$. The QD-cavity mode coupling strength ($g$) is also found to be an efficient handle to manipulate the all optical switching behaviour.

\begin{figure}[h]
\hspace{-0.0cm}
\includegraphics [scale=0.60]{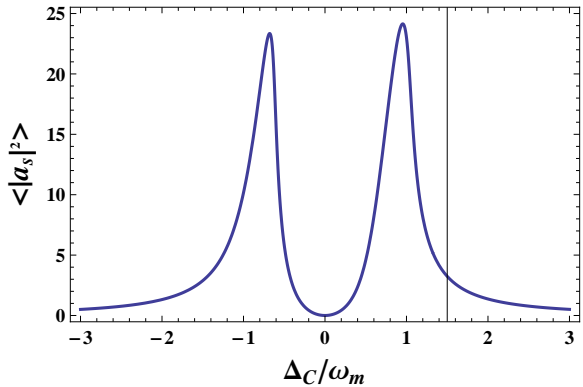}
\caption{Plot of mean intracavity photon number $|a_{s}|^{2}$ versus $\Delta_{c}/\omega_{m}$.Other system parameters used are $\Delta_{d}=\Delta_{c}$, $\eta=0.4 \omega_{m}$, $E_{l}=2 \omega_{m}$, $\gamma_{a}=0.01 \omega_{m}$, $\kappa=0.01 \omega_{m}$, $\chi=0.04$ and $g=0.8 \omega_{m}$. }
\label{Fig.5}
\end{figure}

\section{Controlled Fano resonance and OMIT}

\begin{figure}[h]
\hspace{-0.0cm}
\begin{tabular}{cc}
\includegraphics [scale=0.55]{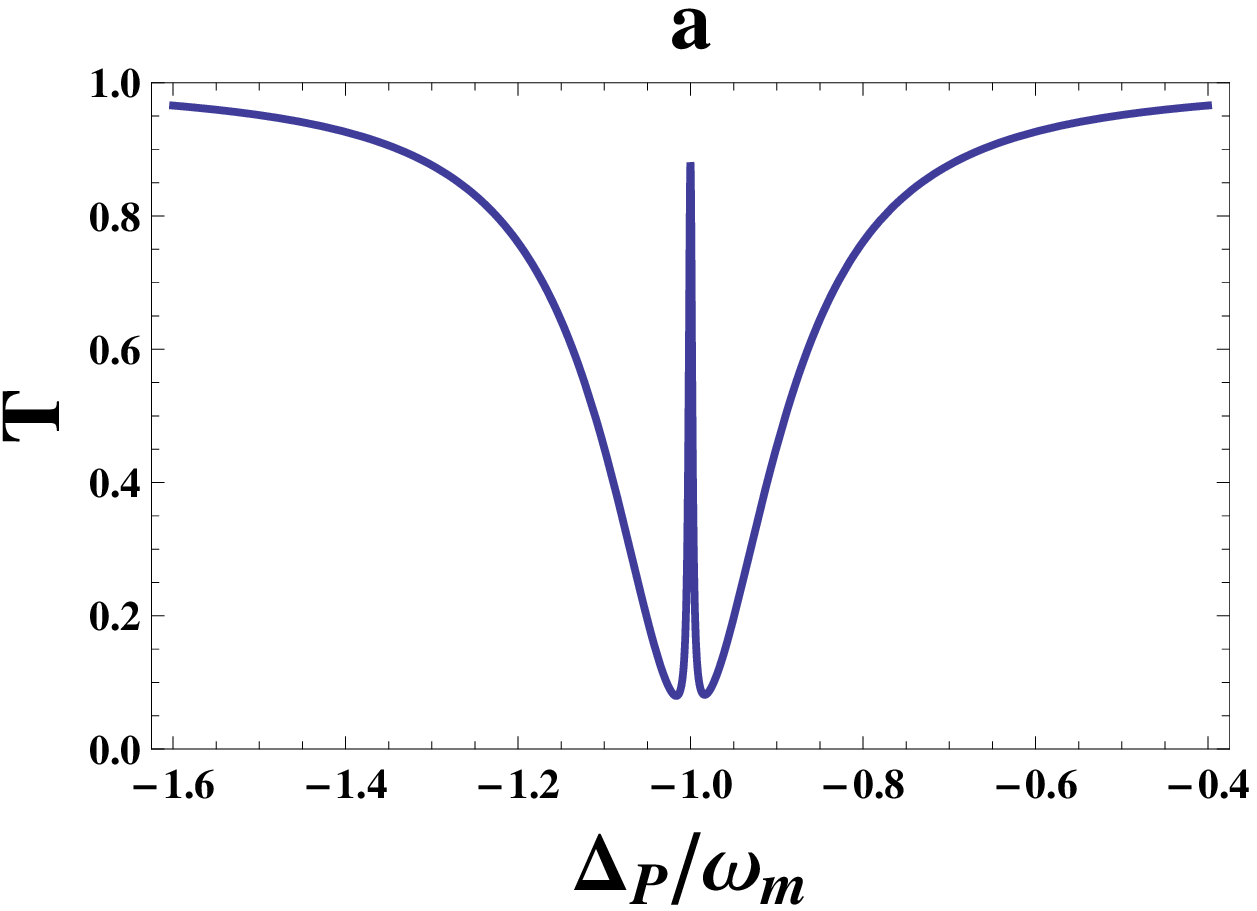}&\includegraphics [scale=0.55] {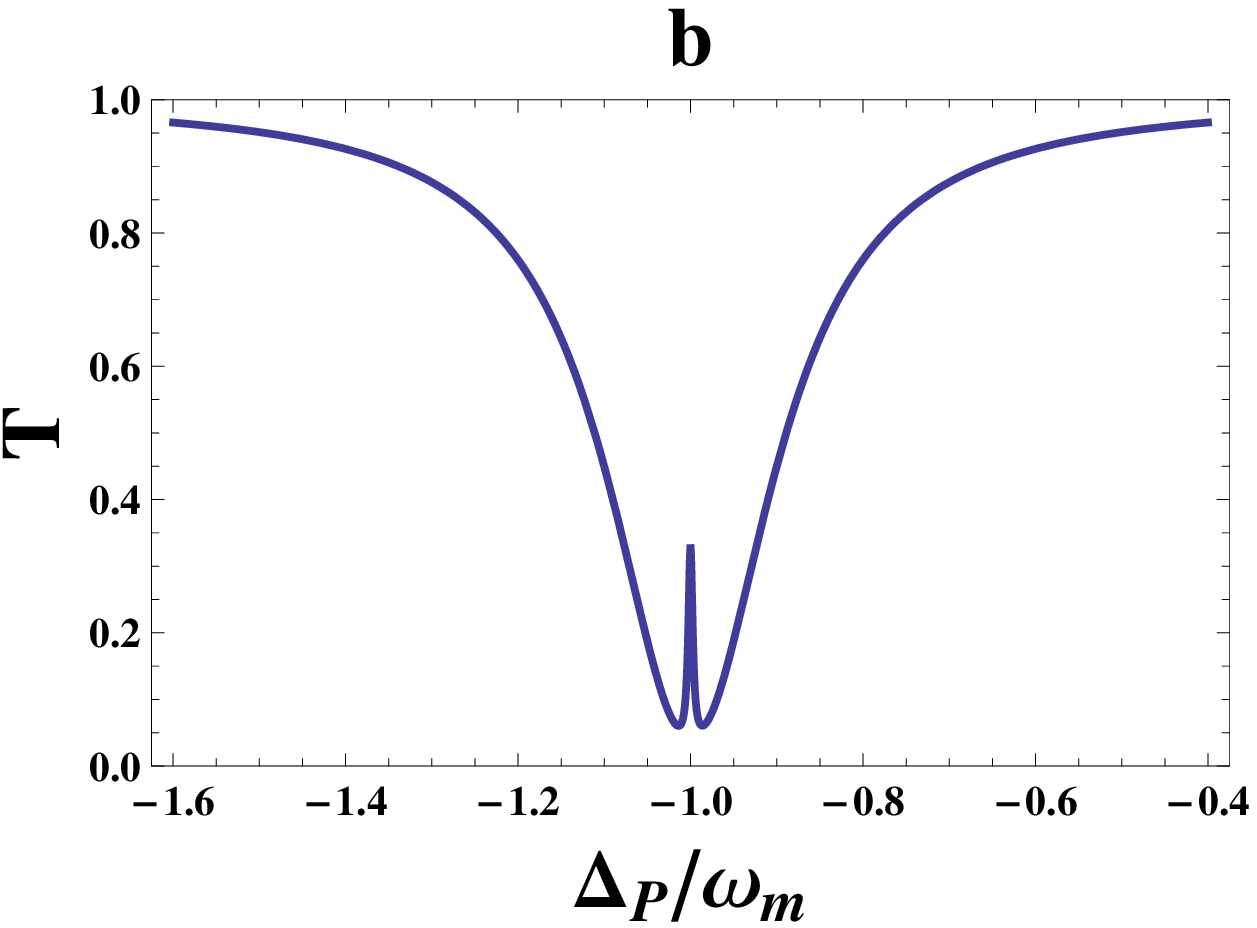}\\
\includegraphics [scale=0.55]{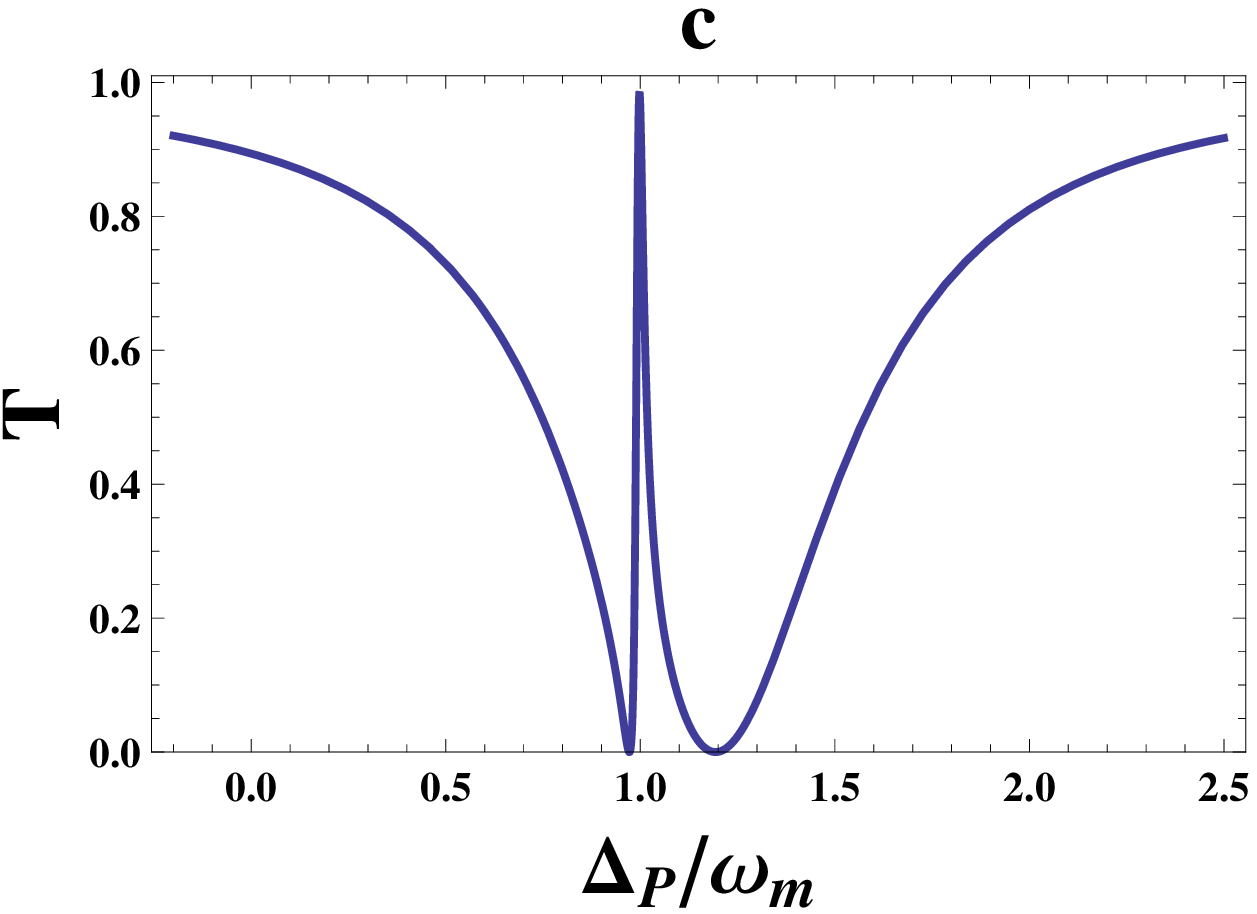}& \includegraphics [scale=0.55] {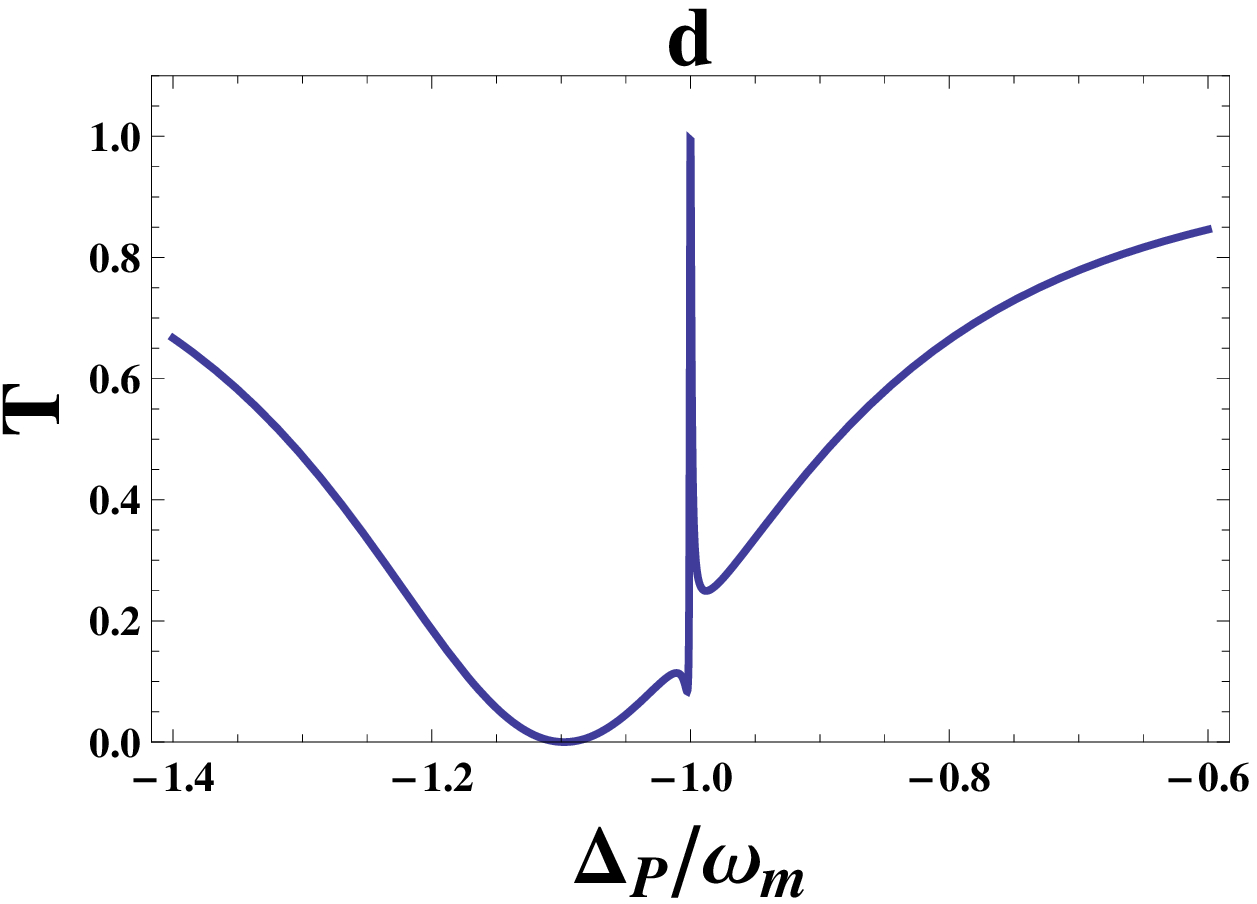}
 \end{tabular}
\caption{The normalized power forward transmission $T$ as a function of the dimensionless detuning $\Delta_{p}/\omega_{m}$. Parameters used are: (a) $\eta=0.113 \omega_{m}$, $\Delta_{c}= -0.9 \omega_{m}$, $\gamma_{m}=0.0017 \omega_{m}$, $\kappa=0.078 \omega_{m}$, $\chi=0.03$ and $a_{s}=\sqrt{0.64}$. (b)$\eta=0.113 \omega_{m}$, $\Delta_{c}= -0.9 \omega_{m}$, $\gamma_{m}=0.017 \omega_{m}$, $\kappa=0.078 \omega_{m}$, $\chi=0.026$ and $a_{s}=\sqrt{0.64}$. (c) $\eta=0.4 \omega_{m}$, $\Delta_{c}= 1.2 \omega_{m}$, $\gamma_{m}=0.001 \omega_{m}$, $\kappa=0.01 \omega_{m}$, $\chi=0.1$ and $a_{s}=1$.(d) $\eta=0.213 \omega_{m}$, $\Delta_{c}= -1.0 \omega_{m}$, $\gamma_{m}=0.0025 \omega_{m}$, $\kappa=0.078 \omega_{m}$, $\chi=0.021$ and $a_{s}=\sqrt{0.64}$.}
\label{Fig.6}
\end{figure}

The OMIT has a prominent sharp symmetric line shape profile which is different from the sharp asymmetric Fano-resonance spectral profile \citep{zha34}. Due to the sharp asymmetric line-shape of Fano-resonance, any small changes in the system parameters can cause a significant change in the amplitude and phase which can actually be used for a sensitive optical switch.
We now explore the effect of the system parameters on controlling and tuning the asymmetric Fano resonance spectral profile. To this end, we will be studying the normalized power forward transmission $T$ which can be measured at the output of the hybrid device. The results discussed in this section are based on the analytical expressions derived in Appendix A. Fig.6 shows the normalized power forward transmission $T$ as a function of the detuning $\Delta_{p}=\omega_{p}-\omega_{l}$ (in the units of $\omega_{m}$). In fig.6(a), we have taken the optomechanical nonlinearity ($\chi=0.03$) to be smaller than the  Kerr nonlinearity ($\kappa=0.0778 \omega_{m}$).It is seen from fig. 6(a), the plot of $T$ has a single transparent peak in the center of $\Delta_{p}=-\omega_{m}$ and two symmetric dips on both sides demonstrating an OMIT effect studied in numerous earlier works \citep{wei27,ma28,aga29,tas30,saf31,ma32,jia33}. In fig 6(b) all parameters are kept same as in fig. 6(a) except the optomechanical coupling strength is reduced $\chi=0.026$. This reduces the transparency peak as seen in fig. 6(b).
In fig. 6(c), the optomechanical nonlinearity ($\chi=0.01$) is taken to be larger than the Kerr nonlinearity ($\kappa=0.01 \omega_{m}$) and plot clearly displays a transparent peak at $\Delta_{p}=\omega_{m}$ with two highly asymmetric dips on either side. This phenomenon shows an obvious asymmetric Fano resonance line-shape. On approaching the Fano resonance ($\Delta_{p}=\omega_{m}$) from the left, the transition from the dip to the transparent peak is extremely sharp while approaching the Fano resonance from the right, the transition from the dip to the peak is slow. In fig 6(d), the system parameters are changed in comparison to figs. 6(a), 6(b) and 6(c), keeping $\kappa$ ($0.078 \omega_{m}$) greater than $\chi$ ($0.021$). We immediately observe an extremely sharp asymmetric Fano peak at $\Delta_{p}=- \omega_{m}$.
In our hybrid optomechanical system we observe that the forward transmission contrast of the Fano resonance is very high (as high as $70 \%$) which meets the minimum criterion for any Telecom system \citep{ree60}. The Fano resonance in our proposed system can be manipulated effectively by the optomechanical coupling $\chi$ and the Kerr nonlinearity $\kappa$.

The underlying physical mechanism for generating such Fano resonances can be understood as follows (based on analytical calculations given in Appendix B). The beat of the pump field and the probe field generates a time-varying radiation pressure force with beat frequency $\Delta_{p}=\omega_{p}-\omega_{l}$. When $\Delta_{p}=\omega_{m}$, the mechanical resonator is driven resonantly. As a result, side-bands of the optical field are generated due to the mechanical oscillations. The frequency of the dominant side-band becomes degenerate with the frequency of the probe field (fig.7). Destructive interference between the side-band and the probe field leads to the cancelation of the intra-cavity field resulting in a transparency window in the transmission.

In view of the above discussions, we find that the QD-cavity coupling, Kerr nonlinearity and the optomechanical coupling plays a key role in the optical switching behavior of the proposed hybrid system. The current system provides a simple and flexible way to tune and coherently control the optical bistability and the Fano resonance spectrum by simply changing the experimentally achievable parameters. The effective optomechanical coupling can be tuned by changing the input laser power, the length of the cavity and the cavity mode frequency. The volume of the cavity and the cavity mode frequency also tunes the Kerr nonlinearity.

The ideas presented in this study could be tested on semiconductor optomechanical systems with built-in quantum dots. Of specific interest are miniature Gallium-Arsenide optomechanical resonators combining strong optomechanical \citep{din61, din62} with cavity QED couplings \citep{pet63}. These hybrid QD-optomechanics platforms are not far from being realized in laboratories and technological advancements will allow to transfer these basic concepts of quantum optomechanics to applications.

\section{Conclusions}

In conclusion, we have analyzed the performance of the optical and opto-mechanical nonlinearities present in a coupled QD-optomechanical cavity system in the presence of a nonlinear Kerr medium. This proposed system can perform an efficient all optical switching. The optical bistability and the sharp asymmetric Fano spectral profile can be controlled and tuned by appropriately changing the QD-cavity coupling, Kerr nonlinearity and the optomechanical coupling. Our results demonstrate the Fano resonance control can be useful for enhancing the sensitivity of optical sensors and designing all optical switches and modulators.

\begin{acknowledgments}
A. B acknowledges financial support from the University Grants Commission, New Delhi under the UGC-Faculty Recharge Programme. M.S.H is grateful to Department of Physics, N.I.T, Calicut and School of Physical Sciences, JNU for providing facilities to carry out this research.
\end{acknowledgments}

\section{Appendix A: Derivation of the normalized power forward transmission}

Our starting point would be the Heisenberg-Langevin equations (3)-(6). We now consider the perturbations generated by the weak probe field and write the optical and mechanical mode as a sum of a steady state mean field term and a quantum fluctuations: $a=a_{s}+\delta a$ and $Q=Q_{s}+\delta Q$. These are substituted in Eqns. (3-6) which yields the results:

\begin{equation}
\psi \delta Q= \omega_{m}^{2} \chi (a_{s}^{*} \delta a+ a_{s} \delta a^{\dagger}),
\end{equation}

\begin{equation}
\frac{d \delta a}{dt}= M \delta a+i \omega_{m} \chi a_{s} \delta Q+ i \kappa a_{s}^{2} \delta a ^{\dagger}+ \sqrt{2 \eta} E_{p} e^{- i\Delta_{p}t},
\end{equation}

where $M=(i \omega_{m} \chi Q_{s}-i \Delta_{c}+ 2i \kappa |a_{s}|^{2}-\eta)$ and $\psi= \frac{d^{2}}{d t^{2}}+\gamma_{m} \frac{d}{dt}+ \omega_{m}^{2}$. The Eqns.(14) and (15) can be further solved by introducing the following ansatz for the fluctuations $\delta a$ and $\delta Q$ as:

\begin{equation}
\delta a = A_{1} e^{-i \Delta_{p}t} + A_{2} e^{i \Delta_{p}t},
\end{equation}

\begin{equation}
\delta Q = Q_{1} e^{-i \Delta_{p}t} + Q_{2} e^{i \Delta_{p}t}.
\end{equation}

Eqns. (16) and (17) are substituted back into Eqns. (14) and (15). This yields the following algebraic equations:

\begin{equation}
Q_{1}= Z(\Delta_{p})\omega_{m}^{2} \chi \left( a_{s}^{*} A_{1}+ a_{s} A_{2}^{*}  \right),
\end{equation}

\begin{eqnarray}
&& ( -i \Delta_{p}-M - i \omega_{m}^{3} \chi^{2} |a_{s}|^{2} Z(\Delta_{p})  )A_{1} \\ \nonumber
&=& a_{s}^{2} ( i \kappa+ i \omega_{m}^{3} \chi^{2} Z(\Delta_{p}) ) A_{2}^{*}+\sqrt{2 \eta} E_{p},
\end{eqnarray}

\begin{eqnarray}
&& ( -i \Delta_{p}-M^{*} + i \omega_{m}^{3} \chi^{2} |a_{s}|^{2} Z(\Delta_{p})   ) A_{2}^{*} \\ \nonumber
&=& a_{s}^{* 2} ( i \kappa+ i \omega_{m}^{3} \chi^{2} Z(\Delta_{p}) ) A_{1},
\end{eqnarray}

where $Z(\Delta_{p})= 1/(\omega_{m}^{2}-\Delta_{p}^{2}-i \gamma_{m} \Delta_{p})$. Upon solving these above equations, we have the expression for $A_{1}$ as

\begin{equation}
A_{1}= \frac{\sqrt{2 \eta} E_{p} X(\Delta_{p})}{\left[ |a_{s}|^{4} (\omega_{m}^{3} \chi^{2} Z(\Delta_{p})+\kappa)^{2}-X(\Delta_{p}) Y(\Delta_{p})   \right]},
\end{equation}

where $X(\Delta_{p})= \left[i \Delta_{p}+M^{*}-i \omega_{m}^{3} \chi^{2} |a_{s}|^{2} Z(\Delta_{p})   \right]$ and $Y(\Delta_{p})= \left[i \Delta_{p}+M+i \omega_{m}^{3} \chi^{2} |a_{s}|^{2} Z(\Delta_{p})   \right]$. Applying the standard input-output theory \citep{walls}, $a_{out}=a_{in}-\sqrt{2 \eta} a(t)$, where $a(t)= E_{l} e^{-i \omega_{l}t}+E_{p} e^{-i \omega_{p}t}$, we have

\begin{eqnarray}
a_{out} &=& (E_{l}-\sqrt{2 \eta} a_{s})e^{-i \omega_{l}t} + (E_{p}-\sqrt{2 \eta}A_{1})e^{-i \omega_{p}t} \\ \nonumber
&-& \sqrt{2 \eta} A_{2} e^{-i (2 \omega_{l}-\omega_{p})t}.
\end{eqnarray}

The expression Eqn.(22) shows the generation of additional frequency components ($2 \omega_{l}-\omega_{p}$) called the Stroke's field. Our interest lies in the transmission of the weak probe field. The normalized power transmission of the weak probe field is given by the expression

\begin{equation}
T= \left| \frac{E_{p}-\sqrt{2 \eta} A_{1}}{E_{p}}  \right|^{2}
\end{equation}

Substituting the expression for $A_{1}$ into Eqn.(23) gives us

\begin{equation}
T= \left| 1- \frac{2 \eta X(\Delta_{p})}{|a_{s}|^{4} (\omega_{m}^{3} \chi^{2} Z(\Delta_{p})+\kappa)^{2}-X(\Delta_{p}) Y(\Delta_{p}) }  \right|^{2}
\end{equation}

\section{Appendix B: Calculations demonstrating the origin of OMIT}

\begin{figure}[htbp]
\centering
\fbox{\includegraphics[width=\linewidth]{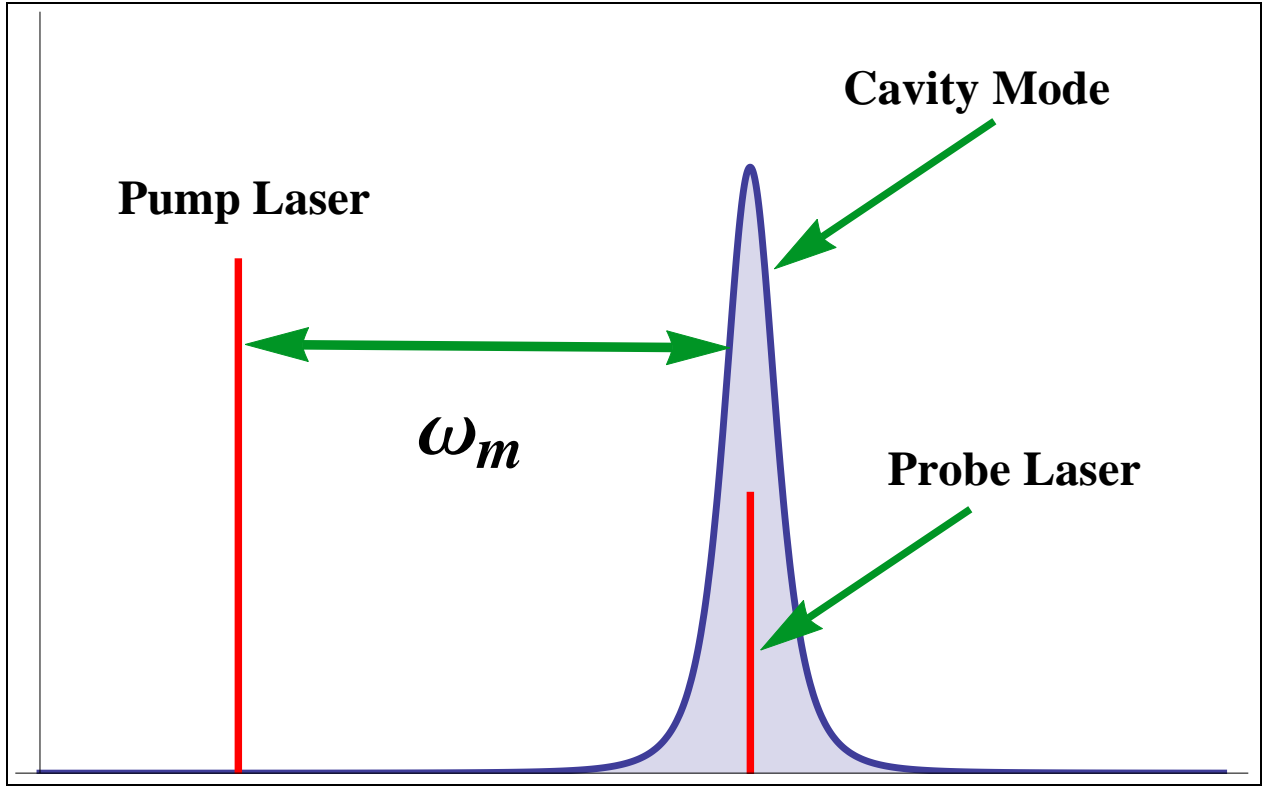}}
\caption{Frequency domain illustration of the pump laser, probe
laser, and the cavity mode. }
\label{Fig.7}
\end{figure}

We start with the fully quantized hamiltonian of our hybrid optomechanical device:

\begin{eqnarray}
H &=& \hbar \omega_{c} a^{\dagger}a + \hbar \omega_{m} b^{\dagger} b + \hbar \omega_{d} \sigma^{\dagger} \sigma - \frac{\hbar}{2} \kappa a^{\dagger} a^{\dagger}a a   \\ \nonumber
&+& \hbar g \left( a^{\dagger} \sigma + \sigma^{\dagger} a \right) -\hbar G a^{\dagger}a (b^{\dagger}+b)  \\ \nonumber
&+&  i \hbar \sqrt{2 \eta}E_{l} \left( a^{\dagger}  e^{-i \omega_{l}t}-a   e^{i \omega_{l}t}\right) \\ \nonumber
&+& i \hbar \sqrt{2 \eta} E_{p} \left( a^{\dagger}  e^{-i \omega_{p}t}-a   e^{i \omega_{p}t}\right).
\end{eqnarray}

Here in addition to the terms already described earlier, we have $b (b^{\dagger})$ as the annihilation (creation) operator for the mechanical mode. As before we obtain the Heisenberg-langevin equations of motion based on the Hamiltonian of Eqn.(25) and solve for the steady states and the quantum fluctuations. The cavity mode fluctuation $\delta a$ is derived as:

\begin{equation}
\delta a= \frac{\sqrt{2 \eta} E_{p}}{i (\Delta_{c}^{'}-\Delta_{p})+\eta+ \frac{|G_{\alpha}|^{2}}{i(\omega_{m}-\Delta_{p})+\gamma_{m}}},
\end{equation}

where $\Delta_{c}^{'}= \Delta_{c}-G (b_{s}+b_{s}^{*})+2 \kappa |a_{s}|^{2}$. Here $G_{\alpha}= G |a_{s}|$ is the coherent intra-cavity field enhanced optomechanical coupling strength.It is clear that the intra-cavity field $\delta a$ of Eqn.26 has the same form as that for a typical optical system exhibiting EIT effect. The pump field plays the role of the coupling between the probe and the mechanical mode.

\end{document}